\documentclass[
    reprint,
    superscriptaddress,
    amsmath,amssymb,
    aps,
    prx,
    longbibliography
]{revtex4-1}

\usepackage{graphicx}
\usepackage{amsmath}
\usepackage{amssymb}
\usepackage[colorlinks=true,urlcolor=blue,linkcolor=blue,citecolor=blue]{hyperref}
\usepackage[margin=0.75in]{geometry}
\usepackage{braket}
\usepackage{xcolor}
\definecolor{LBlue}{rgb}{0,0.34,0.45}

\newcounter{edit}
\newcommand{\edit}[1]{
  \addtocounter{edit}{1}
  \textcolor{red}{[C\theedit] #1}
}

\renewcommand{\edit}[1]{#1}

\begin{document}

\title{Large scale quantum chemistry with tensor processing units}

\author{Ryan Pederson}
\email{pedersor@uci.edu}
\affiliation{Department of Physics and Astronomy, University of California, Irvine, CA 92617, USA}
\affiliation{X, the Moonshot Factory, Mountain View, CA 94043, USA}
\affiliation{Sandbox@Alphabet, Mountain View, CA 94043, USA}

\author{John Kozlowski}
\affiliation{Department of Chemistry, University of California, Irvine, CA 92617, USA}
\affiliation{X, the Moonshot Factory, Mountain View, CA 94043, USA}
\affiliation{Sandbox@Alphabet, Mountain View, CA 94043, USA}

\author{Ruyi Song}
\affiliation{Department of Chemistry, Duke University, Durham, NC 27708, USA}
\affiliation{X, the Moonshot Factory, Mountain View, CA 94043, USA}
\affiliation{Sandbox@Alphabet, Mountain View, CA 94043, USA}

\author{Jackson Beall}
\affiliation{SandboxAQ, Palo Alto, CA, USA}
\affiliation{Sandbox@Alphabet, Mountain View, CA 94043, USA}

\author{Martin Ganahl}
\affiliation{SandboxAQ, Palo Alto, CA, USA}
\affiliation{Sandbox@Alphabet, Mountain View, CA 94043, USA}

\author{Markus Hauru}
\affiliation{The Alan Turing Institute, 96 Euston Road, London, England, NW1 2DB, UK}
\affiliation{Sandbox@Alphabet, Mountain View, CA 94043, USA}

\author{Adam G.M. Lewis}
\affiliation{SandboxAQ, Palo Alto, CA, USA}
\affiliation{Sandbox@Alphabet, Mountain View, CA 94043, USA}

\author{Yi Yao}
\affiliation{Thomas Lord Department of Mechanical Engineering and Materials Science, Duke
University, Durham, NC 27708, USA}

\author{Shrestha Basu Mallick}
\affiliation{X, the Moonshot Factory, Mountain View, CA 94043, USA}
\affiliation{Sandbox@Alphabet, Mountain View, CA 94043, USA}

\author{Volker Blum}
\affiliation{Department of Chemistry, Duke University, Durham, NC 27708, USA}
\affiliation{Thomas Lord Department of Mechanical Engineering and Materials Science, Duke
University, Durham, NC 27708, USA}

\author{Guifre Vidal}
\affiliation{X, the Moonshot Factory, Mountain View, CA 94043, USA}
\affiliation{Sandbox@Alphabet, Mountain View, CA 94043, USA}
\affiliation{Google Quantum AI, Mountain View, CA 94043, USA}
%\date{\today}

\begin{abstract}
We demonstrate the use of Google’s cloud-based Tensor Processing Units (TPUs) to accelerate and scale up conventional (cubic-scaling) density functional theory (DFT) calculations. Utilizing 512 TPU cores, we accomplish the largest such DFT computation to date, with 247848 orbitals, corresponding to a cluster of 10327 water molecules with 103270 electrons, all treated explicitly. Our work thus paves the way towards accessible and systematic use of conventional DFT, free of any system-specific constraints, at unprecedented scales.
\end{abstract}

\maketitle

\section*{introduction}

Computational methods for quantum chemistry and quantum physics have proven to be invaluable tools in modern scientific research and technological innovation. The application space of such methods is vast, ranging from the prediction of novel high-temperature superconductors~\cite{duan2019ab} to the acceleration of drug discovery~\cite{cavasotto2018quantum}; from the study of catalytic processes for e.g. CO$_2$ sequestration~\cite{yu2017co2} and plastic recycling~\cite{jones2016computational} to the design of nanomaterials~\cite{boschetto2021exploring}, solar cells~\cite{frost2014atomistic}, and batteries~\cite{urban2016computational}.   

In the landscape of quantum-based computational methods, density functional theory (DFT) especially stands out due to its ability to produce accurate results for a wide range of systems at a relatively low computational cost~\cite{jones2015density}. Accordingly, an impressive amount of computational research utilizes DFT calculations each year. For instance, the US National Energy Research Scientific Computing Center (NERSC) reported that nearly 30\% of their supercomputer resources in 2018 were spent on DFT calculations alone~\cite{nersc10}. Widespread research and development effort is continuously devoted towards optimizing the performance and accuracy of DFT calculations, giving rise to a plethora of open-source and commercial DFT software packages~\cite{talirz2021trends}. Several packages can leverage specialized hardware, such as general-purpose graphics processing units (GPUs), for most of the workload~\cite{huhn2020gpu,seritan2021terachem,nwchem,gaussian,abinit,bigdft,vasp_gpu}.
However, in conventional DFT implementations, i.e., without specific sparsity assumptions for the density matrix or Hamiltonian matrix, the computational cost scales as the third power of the number $N$ of orbitals used to describe the system (referred to $O(N^3)$ DFT throughout this work), and this cubic scaling often makes simulating large systems, such as protein-ligand complexes or metal-organic frameworks~\cite{wu2017metal}, prohibitively expensive.

\addtocounter{edit}{-1}

\begin{figure}%[ht]
\includegraphics[width=.45\textwidth]{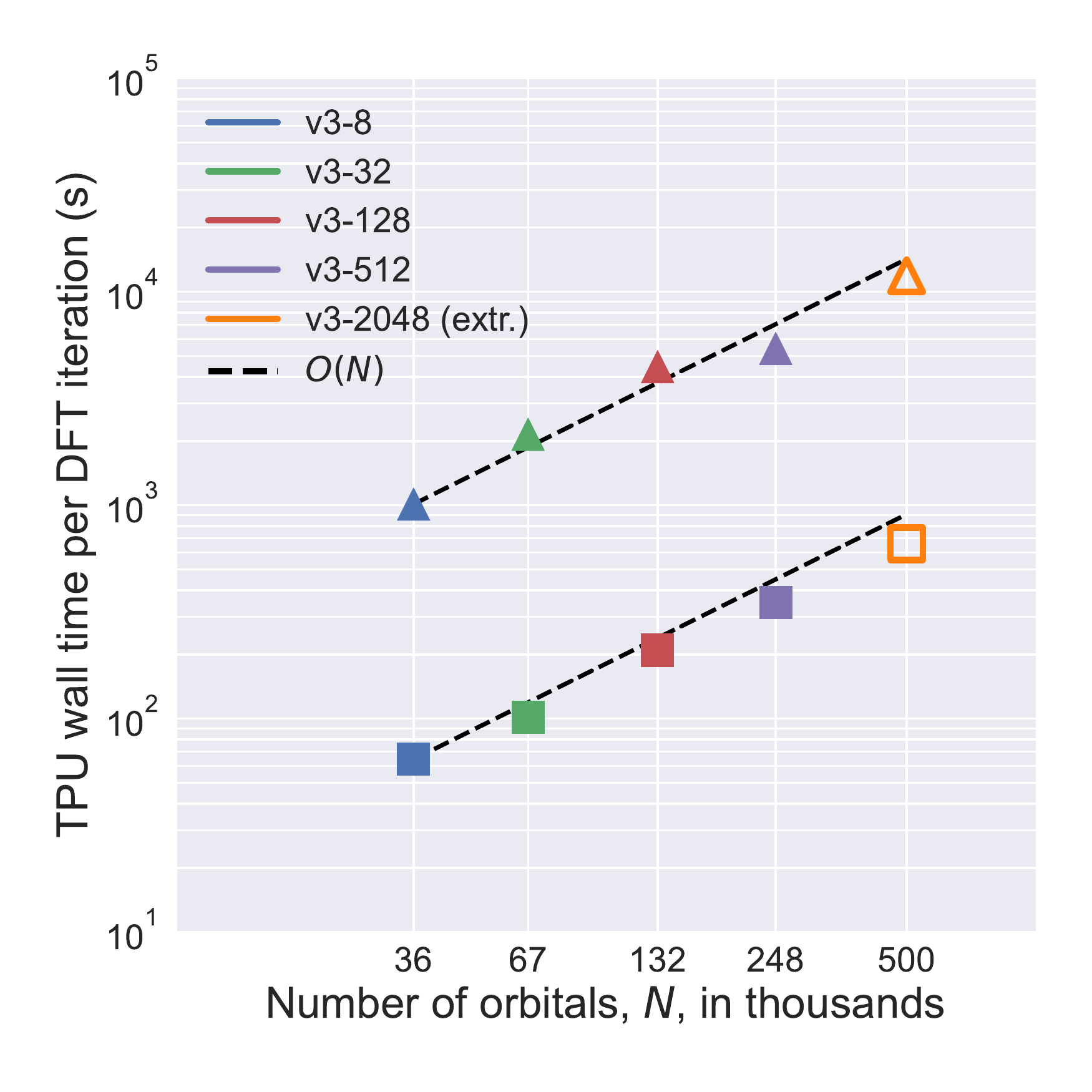}
\centering
\caption{TPU v3 wall time for $O(N^3)$ density matrix purification, Eqs. \eqref{eq:transfH}-\eqref{eq:DD}, as a function of the number $N$ of orbitals, for clusters of water molecules, both in single (squares) and double (triangles) precision. A full TPU v3 pod with 2048 cores \edit{and 32 TB of memory is expected to handle $N \sim 500\,000$ orbitals in our current implementation (extrapolated results necessitated by temporary resource unavailability)}.%, as confirmed at an early stage of the project.
%whereas a TPU v4 pod with 8192 cores can handle up to $N = 1\,000\,000$ orbitals (not plotted). 
%from a linear regression over the smaller TPU configurations, due to temporary resource constraints.
} 
\label{fig: multihost_benchmark}
\end{figure}

\addtocounter{edit}{-1}
\begin{table*}[t]
\setlength{\tabcolsep}{0.5em}
\centering
    \begin{tabular}{|r|r|r|c|r|r|c|}
    \hline
    \multicolumn{1}{|c|}{Number of} & \multicolumn{1}{|c|}{Number of} & \multicolumn{1}{|c|}{Number of} & \multicolumn{1}{|c|}{TPU} & \multicolumn{2}{c|}{TPU wall time (s)} &
    \multicolumn{1}{|c|}{Relative energy}\\
    \cline{5-6}
    \multicolumn{1}{|c|}{orbitals} & \multicolumn{1}{|c|}{atoms} & \multicolumn{1}{|c|}{electrons} & \multicolumn{1}{|c|}{configuration} & ~~FP32~~ & ~~FP64~~ & \multicolumn{1}{|c|}{per molecule (mHa)} \\
    \hline
    35\,544~~ &             4\,443~~~ &                14\,810~~ &              v3-8 &               65~~ &             1\,012~~ & 0.934 \\
    65\,668~~ &             8\,211~~~ &                27\,370~~ &             v3-32 &              102~~ &             2\,150~~ & 0.531 \\
    131\,544~~ &            16\,443~~~ &                54\,810~~ &            v3-128 &              209~~ &             4\,465~~ & 0.291\\
    247\,848~~ &            30\,981~~~ &               103\,270~~ &            v3-512 &              350~~ &             5\,434~~ & 0 \\
    \hline
    \end{tabular}
\caption{Tabulated results in Fig.~\ref{fig: multihost_benchmark}, including also number of atoms and electrons. Wall times for the matrix purification step are shown both for single (FP32) and double (FP64) precision. \edit{Energies are relative to the largest calculation, $E[(\text{H}_2\text{O})_{N_{\text{mol}}}]/N_{\text{mol}} - E[(\text{H}_2\text{O})_{10327}]/10327$, where $N_{\text{mol}}$ is the total number of water molecules.} In this sequence, we used a number of TPU cores that grows roughly as $N^2$. As a result, walltimes are seen to roughly scale linearly in $N$, instead of the expected $O(N^3)$ scaling.}
\label{tab: multihost table}
\end{table*}

Google's Tensor Processing Units (TPUs) are application-specific integrated circuits originally designed to accelerate large-scale machine learning workloads ~\cite{TPUinfo, jouppi2017datacenter, jax, Frostig-Leary2018, Abadi-Zheng2016}. By leveraging the JAX library ~\cite{jax, Frostig-Leary2018, Abadi-Zheng2016}, it is nevertheless possible to repurpose TPUs for other computational tasks ~\cite{Belletti-Anderson2020, Wang-Anderson2021, Pan-Mishra2021, Lu-Ma2020, TPUFFT1, TPUFFT2, huot2019highresolution, morningstar2022simulation, tpu_qphys, tpu_algebra, tpu_Z2field, ganahl2022density}. In this work, we demonstrate the use of TPUs as quantum chemistry supercomputers by accelerating the $O(N^3)$ computational bottleneck of DFT approaches which use an auxiliary single-particle kinetic energy approximation, such as Kohn-Sham (KS)~\cite{hohenberg1964inhomogeneous,kohn1965self} and generalized KS (gKS)~\cite{seidl1996generalized} DFT, where gKS admits hybrid DFT functionals. This enables the systematic study of quantum chemistry problems at unprecedented scales. As a concrete demonstration, we performed end-to-end $O(N^3)$ DFT calculations on large clusters of water molecules, reaching a total of $N = 247\,848$ DFT orbitals, corresponding to $10\,327$ water molecules with $103\,270$ electrons, see Fig.~\ref{fig: multihost_benchmark} and Table~\ref{tab: multihost table}. To our knowledge, this is the largest $O(N^3)$ DFT calculation to date, with the previously largest computation consisting of a single $O(N^3)$ DFT iteration with $N\approx 230\,000$ orbitals on Fujitsu's K computer~\cite{hasegawa2014performance}.

% linear-scaling DFT
Some variants of DFT, most notably linear-scaling DFT~\cite{prentice2020onetep, bowler2010calculations, soler2002siesta, vandevondele2005quickstep}, avoid the $O(N^3)$ bottleneck altogether and can thus reach an even larger number of orbitals. However, these variants rely on additional approximations and conditions, such as truncating density matrix elements ~\cite{cole2010protein}, or on special properties of only a subset of density functionals (such as semilocal density functional approximations). In turn, this results in restricted applicability, with e.g. linear-scaling DFT being suitable for insulating systems but not for metals or systems with a small energy gap~\cite{prentice2020onetep}. In practice, conventional $O(N^3)$ DFT is a more preferable choice since it alleviates technical complexity and problem space restrictions associated with current lower-scaling methods, greatly extending the domain of problems to which DFT can be applied reliably and with relative ease.

%or assuming , that limit the range of applicability, failing e.g. for metals and systems with a small energy gap. Density matrix truncations can introduce errors in the total energy that are similar in magnitude to the differences in the ligand-binding enthalpies found in realistic drug interactions~\cite{mardirossian2020novel}, making $O(N^3)$ DFT as considered here a much more preferable choice. Routine use of $O(N^3)$ DFT for practical calculations at affordable cost would alleviate technical complexity and problem space restrictions associated with current lower-scaling methods, greatly extending the domain of problems to which DFT can be applied with relative ease. Linear-scaling DFT is nevertheless a powerful method for problems that are amenable to it, and might be receptive to TPU acceleration (work in progress).

\begin{figure}[tbp]
\centering
\includegraphics[width=1.0\linewidth]{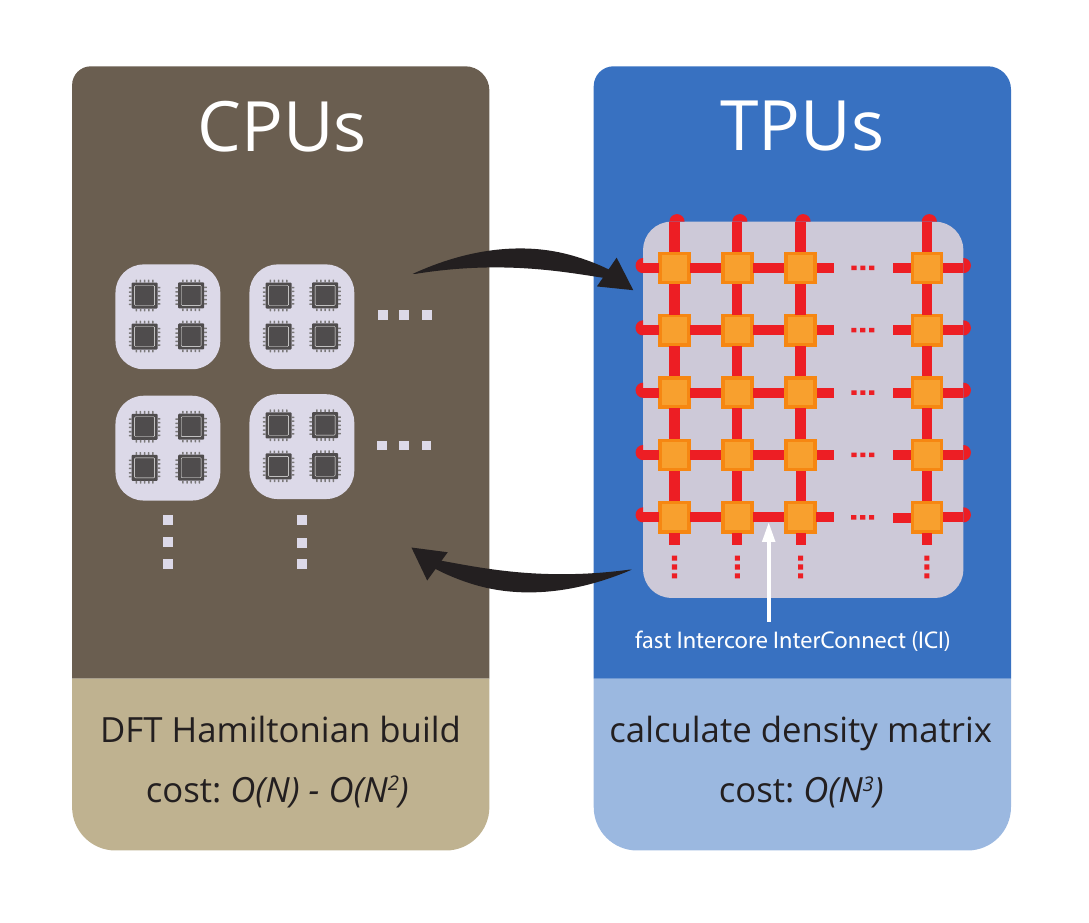}
\caption{The two main steps of our implementation of an $O(N^3)$ DFT computation, the Hamiltonian build and computing the ground state density matrix, which we run on CPUs and TPUs, respectively. The DFT code FHI-aims~\cite{BLUM20092175, FHI} is used to set up the Hamiltonian and the ELSI library~\cite{yu2020elsi,yu2018elsi} is used to facilitate the integration of the TPU-based solver to FHI-aims.
\label{fig:cpu_tpu_graphic}
}
\end{figure}

There are many aspects that go into an $O(N^3)$ DFT calculation. Throughout we focus on atom-centered basis sets with all electrons treated explicitly, that is we do not consider e.g. plane waves or pseudopotentials. At a high level, one can identify two main computational steps: (a) building the DFT Hamiltonian matrix (with cost $O(N)$ to $O(N^2)$) and (b) computing the ground state density matrix ($O(N^3)$), see Fig.~\ref{fig:cpu_tpu_graphic}.

(a) \textit{DFT Hamiltonian build}: Given a choice of $N$ atom-centered basis functions 
%$\left\{ \psi_i(\vec{r}) \right\}_{i=1,\cdots, N}$, 
$\chi_i(\mathbf{r})$, 
one needs to compute the DFT Hamiltonian matrix $H$ and the overlap matrix $S$, with coefficients given by
\begin{equation}
    H_{ij} = \bra{\chi_i} \mathcal{H} \ket{\chi_j}, ~~~~~~~~ S_{ij} = \braket{\chi_i| \chi_j},
\label{eq: H and S}
\end{equation}
where $\mathcal{H}$ represents the DFT Hamiltonian in the continuum and each matrix coefficient requires computing one or several integrals.
%There are many different approaches to building the $N\times N$ Fock matrix $H$ and 
Over the past few decades much effort has been devoted to optimizing the build of the $N\times N$ matrix $H$. Naively, the computational time here scales as $O(N^4)$, however, in many implementations the scaling is effectively reduced to $O(N^2)$ due to two-electron integral screening methods. The scaling can be further reduced to almost $O(N)$ if other strategies, such as fast multipole methods~\cite{white1996linear} or fast fourier transform %(FFT) 
based methods~\cite{vandevondele2005quickstep}, such as the Ewald method for periodic systems, are employed. In this work we do~\emph{not} attempt to accelerate the Hamiltonian or overlap matrix build times with TPUs. Instead, we simply use a well-established all-electron DFT package, the \textit{Fritz Haber Institute ab initio molecular simulations package} (FHI-aims)~\cite{huhn2020gpu,BLUM20092175,FHI}, which we run using CPUs. 

%Note that in DFT implementations with plane waves and other systematic basis sets, the $H$ matrix is usually never built explicitly in practice. Eq.~\eqref{eq: H and S} is only typically executed for basis sets of limited size, compared to the number of orbitals sought. However, for instance in plane wave implementations, the computational scaling remains $O(N^3)$ due to the need for mutual orthogonalization in practice. 

(b) \textit{Density matrix purification:} The pair of matrices $H$ and $S$ define a generalized eigenvalue problem, the so-called KS equations,
\begin{equation} \label{eq:generalized}
H \ket{\phi_\alpha} =e_{\alpha} S \ket{\phi_\alpha},
\end{equation}
with $\ket{\phi_{\alpha}}$ and $e_{\alpha}$ the KS orbitals and energies. Our goal is to compute the ground state density matrix
\begin{equation}
    D \equiv \sum_{\alpha=1}^N \theta(\mu-e_{\alpha}) \ket{\phi_{\alpha}}\bra{\phi_{\alpha}},
\end{equation}
where $\theta(x)$ is the step function and $\mu$ is the chemical potential, chosen such that $\sum_{\alpha=1}^N \theta(\mu-e_{\alpha}) = N_e$, for $N_e$ the number of electrons in the system. The density matrix $D$ can be obtained by solving the KS equations \eqref{eq:generalized} using standard linear algebra libraries, such as LAPACK~\cite{laug} or Intel MKL~\cite{intel-alt}. An alternative route, which we follow in this work, is to use a density matrix purification scheme~\cite{hole_particle_purify, purify_review}. First, by computing the inverse square root of $S$,
\begin{equation} \label{eq:invSquareRoot}
    S \mapsto S^{-\frac{1}{2}}
\end{equation}
we can write the Hamiltonian in an orthonormal basis,
\begin{equation} \label{eq:transfH}
    H \mapsto \tilde{H} \equiv S^{-\frac{1}{2}} H S^{-\frac{1}{2}}.
\end{equation}
and re-express \eqref{eq:generalized} as a standard eigenvalue problem $\tilde{H} \ket{\tilde{\phi}_\alpha} =e_{\alpha}\ket{\tilde{\phi}_\alpha}$,
where $\ket{\tilde{\phi}_{\alpha}} \equiv S^{1/2}\ket{\phi_{\alpha}}$. Next we compute the density matrix $\tilde{D}$,
\begin{equation} \label{eq:purify}
    \tilde{H} \rightarrow \tilde{D} \equiv \theta(\mu I - \tilde{H})= \sum_{\alpha=1}^{N_e} \theta(\mu - e_{\alpha}) \ket{\tilde{\phi}_{\alpha}}\bra{\tilde{\phi}_{\alpha}},
\end{equation}
and finally re-express it in the original basis,
\begin{equation} \label{eq:DD}
    \tilde{D} \mapsto D \equiv  S^{-\frac{1}{2}} \tilde{D} S^{-\frac{1}{2}}.
\end{equation}
\edit{The transformation in Eq.~\eqref{eq:purify} is obtained using a standard density matrix purification scheme that is suitable for TPUs, namely the hole-particle canonical purification scheme~\cite{hole_particle_purify}, which we elaborate on later in the paper.}

If no further modifications are made (e.g., density matrix truncations in linear-scaling DFT), then the cost of computing $D$, whether by solving Eq. \eqref{eq:generalized} or performing the four matrix transformations in Eqs. \eqref{eq:invSquareRoot}-\eqref{eq:DD}, scales as $O(N^3)$. This constitutes what is known as the \textit{cubic wall} of DFT.

The density matrix $D$ is used to derive several important quantities. The real-space electron density $n({\bf r})$ is given by
\begin{equation}
    n({\bf r}) = \sum_{i, j}^N \chi_i({\bf r}) D_{ij} \chi_j({\bf r}) \, ,
\end{equation}
which can be computed on a real-space grid ~\cite{BLUM20092175}.
%using FHI-aims on CPUs~\cite{BLUM20092175}. 
The sum of occupied KS eigenvalues, given by $\mathrm{Tr}(H \, D)=\mathrm{Tr}(\tilde{H} \, \tilde{D})$, is also used to compute the total ground-state energy.
%is computed efficiently on TPUs and is subsequently used in FHI-aims to compute the total ground-state energy. 
Additionally, the~\emph{energy weighted density matrix}~$Q$,
\begin{equation} \label{eq: energy weighted density matrix}
    Q = D \, H \, D \, ,
\end{equation}
is also useful to compute atomic forces analytically~\cite{BLUM20092175}.

\section*{Results}
\subsection*{DFT with TPUs} 
The main result of our work is the successful use of TPUs to perform the four matrix transformations \eqref{eq:invSquareRoot}-\eqref{eq:DD}, thereby tackling the $O(N^3)$ computational bottleneck of DFT. We employed TPUs of the third generation, denoted v3. A single TPU v3 core contains two matrix multiply units (MXUs) to formidably accelerate matrix-matrix multiplication (matmul), resulting in about 10 teraFLOPS (floating point operations per second) of measured single-core matmul performance in single precision. \edit{Importantly for our purposes, matmuls are also available in double precision using a software-emulated 57-bit floating point format. In this approach, utilized algorithms require many more single precision floating point operations when operating in our emulated double precision than in single precision, and as a result matmuls in double precision take $\sim 11\times$ longer than in single precision.} 

%Finally, each TPU v3 core has 16 GB of dedicated high bandwidth memory (HBM). 

The smallest available TPU configuration consists of 8 TPU v3 cores with a total of 128 GB of dedicated high bandwidth memory (HBM), controlled by a single host with 48 CPU cores. The largest configuration is a pod with 2048 TPU v3 cores and 32 TB of HBM, controlled by 256 hosts. 
Given a choice of configuration, the available TPU cores are directly connected to nearest neighbors in a 2D torus network through fast inter-core interconnects (ICIs), see Fig.~\ref{fig:cpu_tpu_graphic}. The ICIs are critical to maintaining high performance when distributing matmuls and other dense linear algebra operations over all available TPU cores. In this work we used the JAX library ~\cite{jax, Frostig-Leary2018, Abadi-Zheng2016} to write \textit{single program multiple data} (SPMD) code and executed it on configurations made of $p$ TPU cores, denoted v3-$p$, for $p=8,32,128$ and $512$. 

\edit{
The TPU hardware architecture is especially suited for dense large-scale matmuls, which we perform in distributed form using the SUMMA algorithm \cite{SUMMA}, as recently demonstrated in Ref. \cite{tpu_algebra}.
Here it was shown that for sufficiently large matrices a v3-512 TPU can perform dense matmuls at near-optimal efficiency: the performance per TPU core (measured in single-precision FLOPS) is maintained at roughly $93\%$ of the single TPU core maximum performance~\cite{tpu_algebra}.}
\edit{It is important to emphasize that TPUs are often ill-suited for other tasks, and hence the algorithms utilized in this work and those in Ref.~\cite{tpu_algebra} had to be picked carefully and may differ from more conventional choices used in CPUs or GPUs.
The use of DM purification algorithms, rather than direct diagonalization, is especially attractive for TPUs since all steps can be evaluated from a series of matmuls.
Clearly, transformations \eqref{eq:transfH} and \eqref{eq:DD} require large-scale matmuls.} Transformations \eqref{eq:invSquareRoot} and \eqref{eq:purify} are implemented by an iteration involving matrix polynomials of small degree, where each polynomial requires a short sequence of matrix additions and multiplications. Specifically, the matrix inverse square root in \eqref{eq:invSquareRoot} is implemented using a standard Newton-Schulz iteration \cite{higham1997stable}, whereas for the density matrix purification in \eqref{eq:purify} we implemented the hole-particle canonical purification scheme~\cite{hole_particle_purify}. Further algorithm details can be found in the Supporting Information.

%\textit{Distributed dense linear algebra on TPUs.---} 
%Recent work \cite{tpu_algebra} demonstrated that TPUs excel at performing a number of distributed dense linear algebra tasks, including matrix-matrix multiplications using the SUMMA algorithm \cite{SUMMA}. In this work, we re-expressed the $O(N^3)$ density matrix purification step of DFT into four operations that can be implemented using mostly matrix-matrix multiplications, which we can then accelerate and scale up by distributing them over the available TPU cores. 

%Specifically, starting from the Fock matrix $H$ and the overlap matrix $S$, we compute (i) the matrix inverse square root of the overlap matrix, $S \mapsto S^{-1/2}$, using a Newton-Schulz iteration that involves small matrix polynomials; (ii) the Fock matrix in an orthonormal basis, $H \mapsto \tilde{H} = S^{-\frac{1}{2}} H S^{-\frac{1}{2}}$, which requires two matrix multiplications; (iii) the purified density matrix $\tilde{D}$ from the Fock matrix $\tilde{H}$, $\tilde{H} \mapsto \tilde{D}$, using the hole-particle canonical purification scheme~\cite{truflandier2016communication}, consisting of a sequence of small matrix polynomials; (iv) the density matrix $D$ in the original basis, $\tilde{D} \mapsto D = S^{-\frac{1}{2}} \tilde{D} S^{-\frac{1}{2}}$, which requires two matrix multiplications. Further details are found in Appendices A and B.

For benchmarking purposes, we have performed end-to-end DFT computations on a sequence of increasingly large water clusters \edit{with geometries obtained from standard molecular dynamics simulations (see Supporting Information)}.  We leverage the DFT code FHI-aims~\cite{BLUM20092175, FHI} to set up and drive calculations using CPUs, then use the TPUs to tackle the $O(N^3)$ dense linear algebra bottlenecks \eqref{eq:invSquareRoot}-\eqref{eq:DD}. We also utilize the \textit{ELectronic Structure Infrastructure} (ELSI) library~\cite{yu2020elsi,yu2018elsi} to facilitate the integration of FHI-aims and the TPU solver. In particular, the DFT Hamiltonian build time and associated computational scaling and parallelization are dictated exclusively by the FHI-aims code,
%and is carried out using CPUs only. 
which uses numeric atom-centered orbitals (NAOs) with an explicit finite spatial extent, and a truncated multipole expansion to accomplish low prefactor and efficient scaling of the Hamiltonian matrix build. 
\edit{While the computational time required to build the DFT Hamiltonian may vary greatly between different systems with the same total number of orbitals $N$ (due to possible differences in the resulting sparsity in the systems), the computational time required for the $O(N^3)$ DM purification step performed on the TPU has much less variability since dense matrix operations are assumed, which do not utilize any sparsity present (see Supporting Information for more discussion). Thus, we emphasize that the TPU wall times, which are reported throughout only for water clusters, are fairly robust with respect to different systems with the same total number of orbitals.}

\begin{figure}[ht]
\includegraphics[width=.45\textwidth]{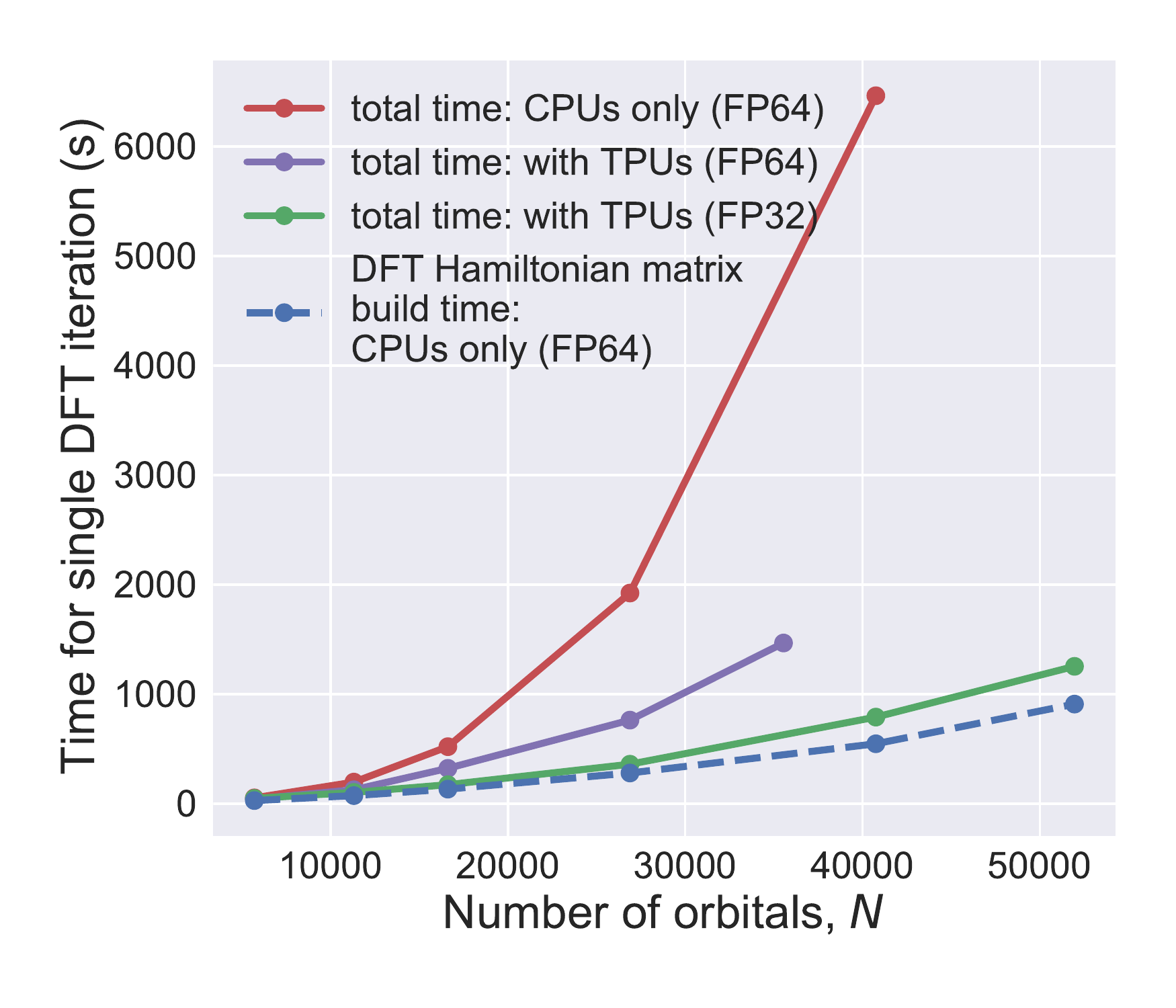}
\centering
\caption{Wall times for a single DFT iteration on water clusters, using a TPU board composed of a CPU host (48 CPU cores) and 8 TPU cores with 128 GB of HBM. 
Green and purple curves correspond to using single and double precision in the TPU solver, respectively. The dashed blue curve corresponds to the CPU time spent on FHI-aims (always in double precision), and should be subtracted from the other curves in order to obtain the time spent on the TPU solvers.  For reference, in red we also plot the time required for a CPU-only computation using the \textit{Eigenvalue soLvers for Petaflop Applications} (ELPA), a highly parallelized eigensolver library \cite{kuus2019optimizations,marek2014elpa}, run on 48 CPU cores.
%In all cases, the average number of purification iterations (\textbf{GV: not defined!)} is $< 25$. Solid curves indicate the total wall time (i.e. Fock matrix build and $O(N^3)$ density matrix purification) while the dashed blue curve represents the Fock matrix $H$ build time accomplished in FHI-aims using 48 CPU cores.
} 
\label{fig: timing_benchmark}
\end{figure}

\edit{Throughout this work we perform all-electron calculations using the PBE exchange-correlation functional and utilize an NAO basis set such that each H$_2$O molecule contributes 10 electrons, represented by 24 orbitals (5 for each hydrogen atom and 14 for the oxygen atom).}
First we consider a single TPU board with 8 TPU v3 cores controlled by a host with 48 CPU cores, and we run FHI-aims on the host. Fig.~\ref{fig: timing_benchmark} shows the wall time for a single DFT iteration (including both Hamiltonian build on CPUs and density matrix purification on TPUs) as a function of the size of the water cluster, which ranges from a few thousand to $N \approx 50\,000$ orbitals. When using the TPU solver in single precision (green curve) we see that
%up to $N \approx 50\,000$ orbitals,
the $O(N)$ Hamiltonian build on 48 CPU cores takes longer than the $O(N^3)$ density matrix purification run on 8 TPU cores, thus shifting the bottleneck. Using the TPU solver in double precision is an order of magnitude slower and saturates the TPU's HBM for $N \approx 36\,000$ orbitals.

Then we consider larger TPU configurations, of up to 512 TPU v3 cores, to perform end-to-end DFT computations on larger clusters, of up to $10\,327$ water molecules (or $N= 247\,848$ orbitals). Fig. \ref{fig: multihost_benchmark} shows the TPU wall time for the $O(N^3)$ density matrix purification for one DFT iteration. These include 350 (5\,434) seconds for a density matrix purification in single (double) precision on the largest cluster, demonstrating feasibility of DFT computations at that scale of a quarter of a million orbitals.

\begin{figure}[ht]
\includegraphics[width=.45\textwidth]{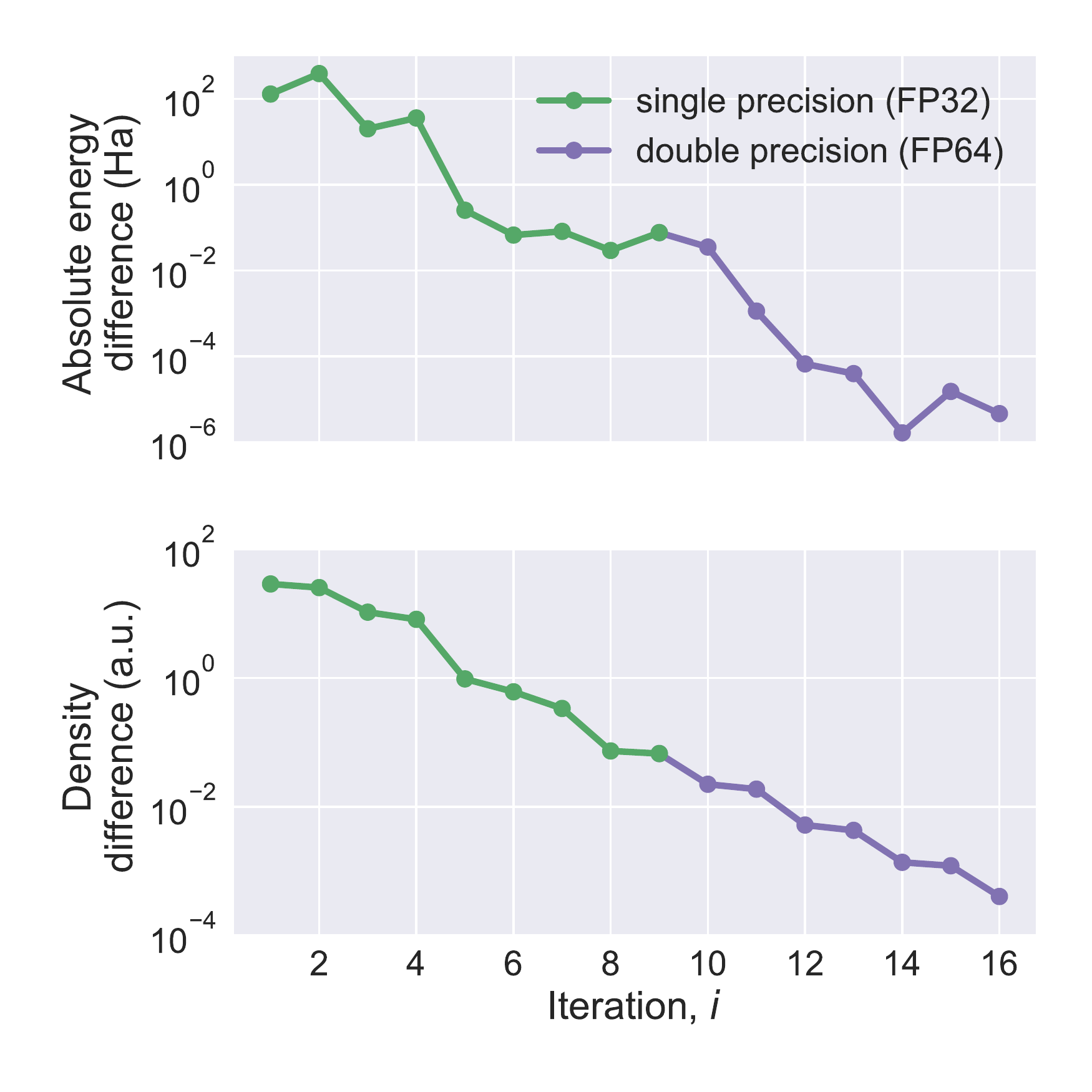}
\centering
\caption{Convergence trajectory of an end-to-end dynamic precision DFT calculation on a (H$_2$O)$_{10327}$ cluster. The absolute total energy differences between subsequent DFT iterations, $i$ and $i-1$, are plotted (top). The corresponding difference in real-space densities within the L$^1$ norm is plotted (bottom).} 
\label{fig:water247848_bifdyn_512_dyn_prec_plot}
\end{figure}

\subsection*{Dynamic precision on TPUs} 

In our implementation, early DFT iterations are treated with single precision and later ones in double precision. This \emph{dynamic precision} approach allows us to cut down on the use of double precision matmuls on TPUs (which are significantly slower than single precision ones) without sacrificing accuracy of the final converged DFT result. Our criteria to switch precision is based on relative density changes, using the L$^1$ norm, defined as $L^1[f(\mathbf{r})] \equiv \int d^3r \, |f(\mathbf{r})| $, and relative energy changes:
\begin{align} \label{eq:converge_density}
    \frac{1}{N_e} L^1[n^{[i]}(\mathbf{r}) - n^{[i-1]}(\mathbf{r})] &< \epsilon \hspace{10mm} \text{and} \, \\
    |E^{[i]} - E^{[i-1]}|/|E^{[i]}| &< \epsilon \, , \label{eq:converge_energy}
\end{align}
where $n^{[i]}(\mathbf{r})$ is the real-space density at DFT iteration $i$, $E^{[i]}$ is the corresponding total ground-state energy, and we use $\epsilon = 5 \times 10^{-7}$ for single precision. 

Fig.~\ref{fig:water247848_bifdyn_512_dyn_prec_plot} shows the convergence trajectory of a dynamic precision DFT calculation for the largest cluster we have considered.
We are able to converge such a DFT calculation to a fairly tight convergence threshold using first $9$ single precision DFT iterations, followed by $7$ double precision DFT iterations. In a smaller cluster, (H$_2$O)$_{1481}$ with $N = 35\,544$ orbitals, a smaller number of double precision iterations are required for convergence, resulting in an overall DFT calculation time that is under $5$ hours on a single TPU board (v3-8), see Supporting Information.

%In addition to single-point DFT energy calculations, analytical forces can be extracted from the TPU-calculated ~\emph{energy-weighted density matrix}, see Appendix~\ref{sec: purification details} (\textbf{GV: is this shown currently?}), enabling large-scale geometry optimization or Ab initio molecular dynamics calculations. 

\section*{Discussion} 

This work has successfully demonstrated that TPUs can both \textit{accelerate} and \textit{scale up} DFT computations. Significant \textit{acceleration} is already achieved using only a single TPU board with 8 TPU v3 cores, see Fig.~3. For instance, an end-to-end dynamic precision DFT computation with N = 35$\,$544 orbitals consisting of 12 iterations in single precision and 4 iterations in double precision yields converged results in under 5 hours. For context, using \edit{double precision only and the highly optimized ELPA $O(N^3)$ solver with 48 CPUs, the same water cluster calculation required 20 hours to achieve $16$ DFT iterations}.

In order to \textit{scale up} the size of DFT computations while retaining high performance two main ingredients are involved: (i) a larger amount of high bandwidth memory, scaling as $O(N^2)$, to be able to store dense $N\times N$ matrices; (ii) a larger number of cores, with state-of-the-art inter-core connectivity, to more effectively execute the $O(N^3)$ floating point operations involved in the required distributed matrix transformations. As shown in Fig.~1, 
%for TPU configurations with $p$=8, 32, 128, 512, and 2048 cores and commensurate amounts of HBM (up to 32 TB), 
by using a number of cores that scales as $O(N^2)$ and commensurate amounts of HBM, we can scale up to $N = 500\,000$ orbitals with wall times that only grow proportional to $N$. 

\edit{Once the main ingredients (i) and (ii) are satisfied, the Input/Output (IO) time, i.e. the end-to-end communication time between the TPU and CPU, can become an important (and possibly limiting) factor. 
This IO step is not exclusive to TPUs, for instance, it is also relevant and analogous in GPU setups that require communication with CPUs.
This step is parallelizable, such that by using a number of TPU and CPU cores that scales as $O(N^2)$, the total IO time remains constant $O(1)$. 
However, in this context, obtaining such optimality will depend on specific implementation details of the DFT code utilized, such as the matrix distribution pattern on the CPU processor grid.
Our current prototype implementation is modular and generalizeable and IO times scale unfavorably as $O(N^2)$, even becoming the rate-limiting step in some cases (see Supporting Information). 
Less-general (but straightforward) engineering approaches are expected to be much more optimal, but are beyond the scope of this work which aims to demonstrate the use of TPUs in a more general context. 
}

We emphasize that other hardware accelerators, most notably GPUs, can also accelerate and scale up DFT computations in a similar manner as discussed above, and Ref.~\cite{das2022dft} recently presented a useful and positive development in this direction.
\edit{
Modern distributed GPU configurations are expected to achieve similar performance to TPUs in this regard, however, a direct comparison is complicated by the highly diverse nature of distributed GPU configurations found in practice. On the Summit supercomputer configuration, it has been reported that 432 distributed Nvidia V100 GPUs can perform matmuls for dense $N > 500\,000$ matrices with a performance per GPU (measured in FLOPS) that is roughly 85\% of the single V100 GPU maximum performance~\cite{herault2019generic}.
}

% beyond water molecules
Here we have focused, for simplicity, on applying DFT to clusters of water molecules. More complicated systems may present additional difficulties. For instance, protein-ligand complexes often require more elaborate schemes, such as including solvation to facilitate the convergence of the DFT iteration~\cite{lever2013electrostatic,rudberg2012difficulties}. Work in progress shows that our TPU-based large-scale DFT computations can also successfully address protein-ligand complexes with explicit solvents, %~\cite{tpu_proteins}
as well as in a variety of other large systems, including DNA segments, carbon nanotubes, and graphene surfaces. In addition to single-point DFT energy calculations, analytical forces can also be extracted from the TPU-calculated ~\emph{energy-weighted density matrix}, enabling large-scale geometry optimization or Ab initio molecular dynamics calculations.

DFT is a highly successful quantum-based method, but it is ultimately a consistent-field approximation, which may not be accurate enough for certain applications. Fortunately, TPUs can also accelerate and scale up other, more accurate quantum chemistry approaches where the computational bottleneck is again given by dense linear algebra operations. For example, in density matrix renormalization group (DMRG)~\cite{white_density_1992} calculations, TPUs can be used to reach an unprecedentedly large bond dimension $D = 65\,536$~\cite{ganahl2022density}. Similarly, we anticipate that TPUs will thrive in other methods such as coupled cluster~\cite{fales2020performance} and Møller–Plesset perturbation theory~\cite{vogt2008accelerating}. Even when applying such higher-level methods, large-scale DFT may still be a crucial piece in simulations that require a quantum-mechanically treated region that embeds a subsystem treated with higher-level correlated methods~\cite{sun2016quantum}.

% summary
To conclude, in this work we have successfully repurposed TPUs as quantum chemistry supercomputers by tackling the $O(N^3)$ computational bottleneck of density functional theory. We demonstrated performance and scalability with a water cluster with $N = 247\,848$ orbitals, which to our knowledge is the largest $O(N^3)$ DFT computation to date. We remark that cloud-based TPUs, and other hardware accelerators such as GPUs, are more accessible and affordable than traditional supercomputer resources. Our work thus paves the way towards accessible and straightforward use of quantum chemistry computational methods for much larger systems than were previously possible. \\

\textbf{Competing Interests}: V.B. received compensation as an advisor from Google during part of this work. V.B. is also a board member of MS1P e.V., the non-profit organization that licenses the FHI-aims electronic structure code used in this work. He does not receive any financial gains from this position.

\begin{acknowledgments}

The authors would like to thank Toru Shiozaki and Garnet Kin-Lic Chan for suggesting to investigate the use of TPUs to accelerate mean-field quantum chemistry methods (by accelerating matrix multiplications for a diagonalization-free construction of the density matrix, as in the density matrix purification used in this paper), and Toru Shiozaki, Garnet Kin-Lic Chan, Chase Roberts and Stefan Leichenauer for previous exploratory work in this direction. Also, special thanks to Xing Zhang and Garnet Kin-Lic Chan for work adjusting PySCF, as part of an on-going integration of our TPU solver, to be described elsewhere, and to David Bowler, Tsuyoshi Miyazaki, and Jun-ichi Iwata for help documenting the largest DFT computation run on the K computer.
Fianlly, the authors would also like to thank 
Giuseppe M. J. Barca, % ANU Australia
Anudhyan Boral,
Michael Brenner,
Kieron Burke, % UC Irvine
Rafael Gomez-Bombarelli, % MIT 
JW Feng,
Filipp Furche, % UC Irvine
Andreas Goeller, % Bayer
Stephan Hoyer,
Olivier Lacombe, 
Stefan Leichenauer,
Lin Lin, % UC Berkeley
Ruben Martin Romo, % ICIQ Barcelona
Todd Martinez,  % Stanford
Anders M. N. Niklasson, % Los Alamos
Nicholas Rubin, % Google Quantum AI
Zak Stone, % Google Research
Matthias Tan, 
Keiran Thompson, % Stanford
Edward Valeev, % Virginia Tech
and
Jae Yoo % 
for useful discussions and comments.
Research supported with Cloud TPUs from Google's TPU Research Cloud (TRC).
Sandbox is a team within the Alphabet family of companies, which includes Google, Verily, Waymo, X, and others. 
G.V. is a CIFAR fellow in the Quantum Information Science Program, a Distinguished Invited Professor at the Institute of Photonic Sciences (ICFO), and a Distinguished Visiting Research Chair at Perimeter Institute. Research at Perimeter Institute is supported by the Government of Canada through the Department of Innovation, Science and Economic Development and by the Province of Ontario through the Ministry of Research, Innovation, and Science. R.S. and Y.Y. were partially supported by the National Science Foundation (NSF), USA under Award No. 1450280.
\end{acknowledgments}

\clearpage

\renewcommand{\thesubsection}{S\arabic{subsection}}
\renewcommand{\theequation}{S\arabic{equation}}
\renewcommand{\thefigure}{S\arabic{figure}}

\setcounter{equation}{0}

\section*{Supporting information}

\subsection{orthogonalization details}
\label{sec: orthogonalization details}

The original set of $N$ basis functions $\chi_i(\mathbf{r})$, corresponding in our current FHI-aims implementation to numeric atom-centered orbitals (NAOs), are not orthogonal, in the sense that the overlap matrix $S$, with coefficients
\begin{equation}
    S_{ij} = \braket{\chi_i| \chi_j} = \int d^3r ~\chi_i(\mathbf{r})\chi_j(\mathbf{r})
\end{equation}
is not the identity matrix, but some non-trivial, positive definite matrix.
To identify the linear combinations of orbitals that are occupied in the ground state of the system, we need to account for overlaps between the orbitals.
This can be done using the L\"{o}wdin decomposition, where the Hamiltonian matrix $H$, with coefficients
\begin{equation}
    H_{ij} = \braket{\chi_i| \mathcal{H}| \chi_j} = \int d^3r ~\chi_i(\mathbf{r}) \mathcal{H}(\mathbf{r})\chi_j(\mathbf{r}),
\end{equation}
is transformed into an orthonormal basis as $H \mapsto \tilde{H} = S^{-\frac{1}{2}} H S^{-\frac{1}{2}}$. The transformed Hamiltonian $\tilde{H}$ can then be purified, as described in the next section, to yield the density matrix $\tilde{D}$, which is then transformed back into the original orbitals basis, 
\begin{equation}
    \tilde{D} \mapsto D = S^{-\frac{1}{2}} \tilde{D} S^{-\frac{1}{2}}.
\label{eq:orthogonal D}
\end{equation}
For this purpose, the inverse square root $S^{-\frac{1}{2}}$ of the overlap matrix is needed.
To compute it efficiently on TPUs, we need an algorithm for computing the matrix inverse square root for which the computational bottleneck reduces to repeated matrix multiplications. As discussed in~\cite{higham1997stable}, this can be achieved using Newton-Schulz iterations. The iteration
\begin{align}
    \label{eq:newton-schulz}
    X_{[n+1]} = \frac{1}{2} X_{[n]} (3I - X_{[n]}^2), \quad X_{[0]} = A,
\end{align}
converges to the matrix sign function of $A$, which turns positive eigenvalues to $+1$ and negative eigenvalues to $-1$. Notice that two matrix multiplications are needed for each iteration. These matrix multiplications can be executed very quickly when distributed over a set of TPUs. The inverse square root can in turn be cast as a sign function as
\begin{align}
    \label{eq:block-signum}
    \operatorname{sgn}\left( \begin{bmatrix} 0 & S \\ I & 0 \end{bmatrix} \right) = \begin{bmatrix} 0 & S^{\frac{1}{2}} \\ S^{-\frac{1}{2}} & 0 \end{bmatrix}.
\end{align}
Applying the iteration \eqref{eq:newton-schulz} to the block matrix in \eqref{eq:block-signum} results in the Denman-Beavers iteration for computing the inverse square root. 

In single (double) precision, the above procedure typically converges in about 35-50 (65-90) iterations, depending on how small the absolute value of the smallest (in absolute value) eigenvalue of matrix $A$ is. In order to further accelerate this computation, we introduce the pre-conditioning polynomial iteration (which we described and justified in Sect. III.D of \cite{tpu_algebra} in the related context of the matrix sign function for singular values),
\begin{align}
    \label{eq:rogue}
    X_{[n+1]} = aX_{[n]} (I - \frac{4}{27} a^2 X_{[n]}^2), \quad X_{[0]} = A,
\end{align}
where $a = \frac{3}{2} \sqrt{3} - s_-$ for some choice of small $s_->0$. [Notice that for $a=3/2$, that is $s_- = 3(\sqrt{3}-1)/2$ we recover \eqref{eq:newton-schulz}.] This pre-conditioning polynomial accelerates the growth of small eigenvalues of $A$, until they become of size at least $s_{-}$. From then on, the regular Newton-Schulz iteration is used to bring all the positive eigenvalues to 1, with quadratic convergence.  For $s_- = 0.1$, in single (double) precision we need 15-20 (35) iterations of the pre-conditioning polynomial and 10 (10) iterations of the regular polynomial, for a total of 25-30 (45) iterations. 

The inverse of $S$ is of course very sensitive to poor conditioning of $S$, which for the overlap matrix corresponds to (nearly) linearly dependent orbitals, a common occurrence when dealing with large molecules.
The danger of instability and poor accuracy due to small eigenvalues of $S$ is especially pressing if operating in low numerical precision.
Consequently we always compute $S^{-\frac{1}{2}}$ in double precision.
Because TPUs do not operate natively in double precision but rather rely on software emulation, this incurs a significant time cost.
However, in a full DFT simulation that cost gets amortized:
the overlap matrix does not change between DFT iterations (only the Hamiltonian does), and thus we only need to compute $S^{-\frac{1}{2}}$ once at the first iteration, write the result to disk, and reread and use it at all the successive iterations with negligible cost.

\subsection{purification details}
\label{sec: purification details}
By \textit{density matrix purification} we mean the map from a Hermitian matrix $\tilde{H}$ of linear size $N$  to a certain density matrix $\tilde{D}$, itself a projector into the subspace corresponding to the $N_e$ smallest (or most negative) eigenvalues of $\tilde{H}$, where $N_e$ is the number of electrons in the system. [As in the rest of the paper, the tilde in the Hamiltonian $\tilde{H}$ and density matrix $\tilde{D}^{(k)}$ denotes that these matrices are expressed in an orthonormalized basis of orbitals, as described in the previous section.] In symbols, let 
\begin{equation}
    \label{eq:Hdecomp}
    \tilde{H} = V \Sigma V^H
\end{equation}
be the ascendingly sorted eigendecomposition of $\tilde{H}$, and let $\rho \equiv \mathrm{diag}(1_1, 1_2, \ldots, 1_{N_e}, 0_{N_e+1}, 0_{N_e+2}, \ldots, 0_N)$ be a diagonal matrix with $N_e$ 1's followed by $N - N_e$ 0's on the main diagonal. Then $\tilde{D}$ is defined by
\begin{equation}
    \label{eq:Ddecomp}
    \tilde{D} \equiv V \rho V^H.
\end{equation}
It follows identically that $\tilde{D}^2 = \tilde{D}$, so that $\tilde{D}$ is indeed a projector.

With the decomposition \eqref{eq:Hdecomp} in hand, $\tilde{D}$ is trivially computed by the manual substitution $\Sigma \to \rho$. Unfortunately an efficient algorithm for Hermitian eigendecomposition is not presently available in a distributed-TPU context. Instead, we turn to matrix-multiplication based purification algorithms originally developed in the context of linear scaling methods (see \cite{purify_review} for a review; note that since our matrices are dense and are not truncated, our implementations scale as $N^3$ despite the name). 

Density matrix purification algorithms can be divided into two classes \cite{canonical_purify} by the manner in which $N_e$ is specified. In \textit{grand canonical purification}, a so-called \textit{chemical potential} $\mu$ is given, and $\tilde{D}$ found so that the $N_e+1$'th most negative entry of $\Sigma$ is the first to exceed $\mu$. This can be achieved by shifting the spectrum of $\tilde{H}$ by $\mu$ so that the latter divides negative from positive eigenvalues, and then computing a polar decomposition using the methods of \cite{tpu_algebra}. 

In the \textit{canonical purification} used in this work, $N_e$ is instead specified directly. Compared to the grand canonical case this is more directly relevant to computations of molecular electronic structure, where $\mu$ is unknown but the number $N_e$ of electrons is provided.

Various algorithms for canonical purification have been proposed in the literature. The original scheme is presented in \cite{canonical_purify}, and is variously referred to as \textit{canonical purification} (in which case other algorithms are given a different name), \textit{trace-preserving purification}, or the \textit{Palser and Manolopoulos scheme}. The \textit{trace-resetting} schemes proposed in \cite{trace_resetting_purify1, trace_resetting_purify2} are probably most common in practical use. We use the \textit{generalized} or \textit{hole-particle} scheme presented in \cite{hole_particle_purify}. In our TPU experiments this iteration yields  performance comparable to that of \cite{trace_resetting_purify1, trace_resetting_purify2}, but avoids certain branching conditionals which are awkward to phrase efficiently on the TPU.

All such schemes work by first mapping the input $\tilde{H}$ to some initial $X_{[0]}$ with eigenvectors unchanged but eigenvalues bound in $[0, 1]$, and then repeatedly applying a matrix-multiplication based iteration which also preserves eigenvectors. This iteration is chosen so that the eigenvalues of its fixed point $X_{[\infty]}$ are \emph{exactly} either 0 or 1 with $\mathrm{Tr}(X_{[\infty]}) = N_e$; $X_{[\infty]}$ then satisfies \eqref{eq:Ddecomp} and it is thus equal to $\tilde{D}$, up to numerical error. \edit{In practice, the number of purification iterations required for convergence varies across different Hamiltonians and the numerical precision desired. Relevant factors include the size of the energy gap and the fraction of occupied to unoccupied states. Typically, less than $50$ purification iterations are required~\cite{hole_particle_purify}. Calculations performed to double precision tend to require more purification iterations, up to twice as many as the same calculation performed to single precision.}

Details of the specific iteration we use are given in \cite{hole_particle_purify}. It can be reproduced by the initialization
\begin{subequations}
\begin{align}
    X_{[0]} &= \beta_1 I + \beta_2(\mu I - \tilde{H}), \\
    \mu &= \frac{\mathrm{Tr}\tilde{H}}{N}, \\
    \beta_1 &= \frac{k / N}{e_{+} - \mu}, \\
    \beta_2 &= \frac{1 - k/N}{\mu - e_{-}},
\end{align}
\end{subequations}
where $e_{+}$ and $e_{-}$ are estimates of the largest and smallest eigenvalues of $\tilde{H}$ obtained by e.g. the Gershgorin circle theorem. Note that in practice we use the slightly more complicated initialization referred to as \textit{HPCP+} in \cite{hole_particle_purify}, which gives moderately improved performance when $N_e$ is far from $N / 2$. In either case, the iterate $X_{[n+1]}$ is found from its predecessor $X_{[n]}$ via
\begin{subequations}
\begin{align}
    X_{[n]}^{\prime} &= I - X_{[n]}, \\
    X_{[n + 1]} &= X_{[n]} + 2 \left(X_{[n]}^2 X_{[n]}^{\prime} - \frac{\mathrm{Tr}(X_{[n]}^2 X_{[n]}^{\prime})}{\mathrm{Tr}(X_{[n]} X_{[n]}^{\prime})} X_{[n]} X_{[n]}^{\prime}\right).
\end{align}
\end{subequations}
Once $\tilde{D}$ is found, its counterpart in the non-orthogonal basis, $D$, is found by applying \eqref{eq:orthogonal D}.

\begin{figure}[ht]
\includegraphics[width=.45\textwidth]{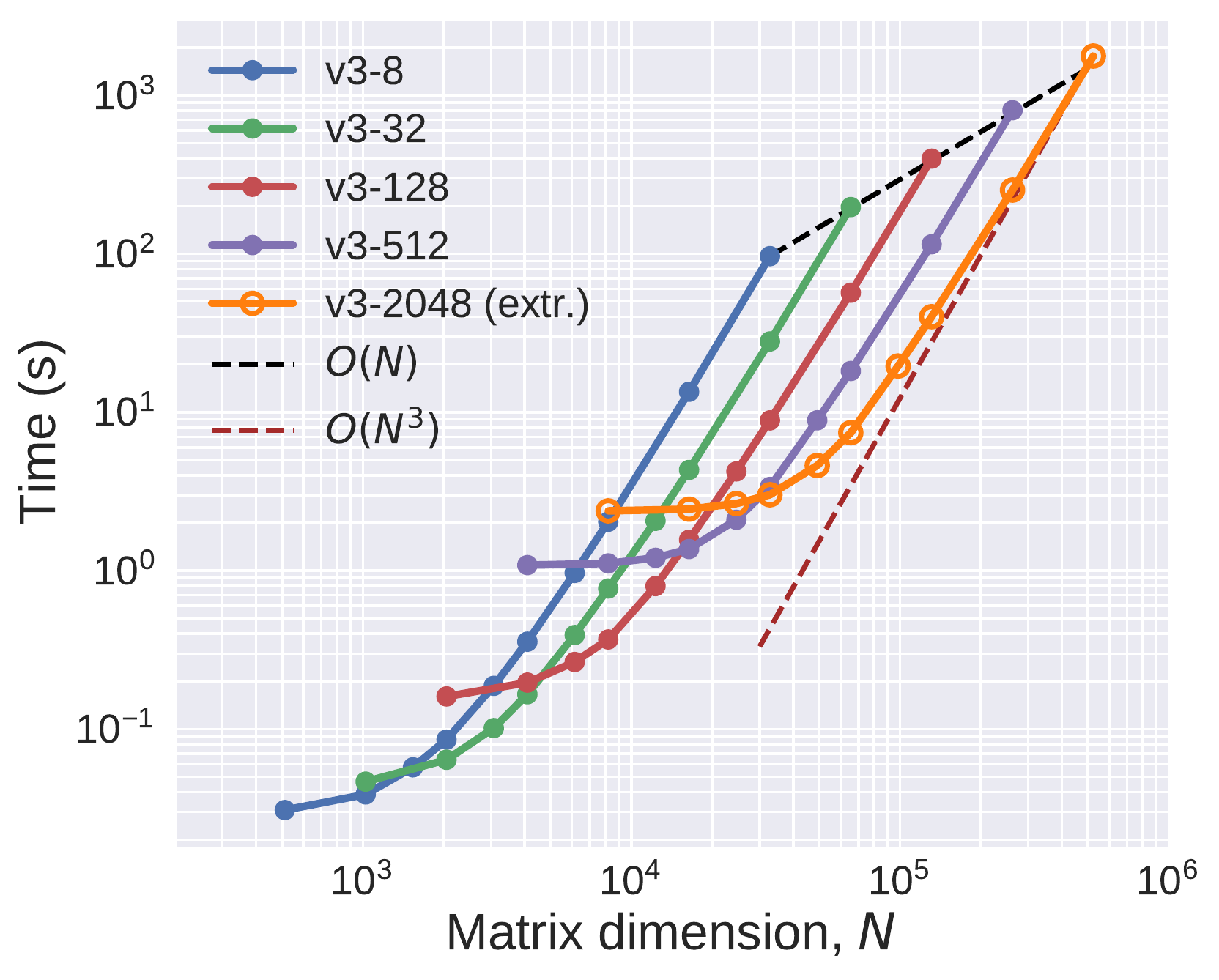}
\centering
\caption{Average wall times for TPU density matrix purification (normalized to a total of $50$ purification iterations) in single precision using dense random Hermitian matrices of dimension $N$. Open-circle data points (v3-2048 results) are a linear extrapolation from v3-512 results.} 
\label{fig: purify_benchmark}
\end{figure}

In Fig.~\ref{fig: purify_benchmark} we demonstrate the computational scaling of density matrix purification on TPUs using dense random Hermitian matrices and single precision. In this benchmark we scale both the system size (dimension of the matrix, $N$) and the number of TPU v3 cores used. Starting with a single TPU board, consisting of 8 TPU v3 cores, we can systematically scale up to hundreds (or thousands) of TPU v3 cores. Using a full TPU v3 pod (consisting of $2048$ TPU v3 cores), we project that we can address dense systems of $N = 500\,000$ orbitals within 30 minutes using single precision.

For dense linear algebra, the computational scaling here is cubic. If suitable sparsity is assumed, and the density matrix is correspondingly truncated, sparse linear algebra can be used to obtain linear scaling. Such linear scaling approaches have been implemented and used in practice within computational quantum chemistry packages, however their practical application is limited to systems whose density matrix has a sufficient sparsity structure to ensure accurate results. 
%  Already in main text
%For example, standard density matrix truncations can introduce errors in the total energy that are similar in scale to the differences in the ligand-binding enthalpies found in realistic drug interactions~\cite{qsimulate paper}.

\subsection{FHI-aims TPU integration details}
\label{sec: fhi-aims}

We outline the practical details of integrating a TPU-based density matrix solver with the CPU-based DFT package FHI-aims. The platform and integration described here is a prototype. 
%and is not anticipated to be performant or made publicly available at this time. 
Its main purpose is to illustrate, in actual end-to-end DFT computations, the viability of accelerating the $O(N^3)$ bottleneck using TPUs. 

The software ELSI~\cite{yu2020elsi,yu2018elsi} is used to facilitate the connection between FHI-aims and the TPU by providing an interface and abstraction in which FHI-aims, or other codes, such as Siesta~\cite{garcia2020siesta} and DFTB+~\cite{hourahine2020dftb+}, can utilize external eigensolvers launched within ELSI. We implement in-house routines to launch the TPU-based density matrix purification (instead of an eigensolver) using the ELSI standard. In all calculations, we use an off-the-shelf FHI-aims code with \emph{no} modifications (version 210226).

% [from Jackson] In Appendix C we discuss the process of moving data between FHI CPU process space and TPUs process space.  We mostly represent our process of moving data as "we serialize and transfer". A few clarifications that might better communicate what we are doing: Mention that each process (cpu and tpu) calculates where its data should be within a serialized CSC representation of the whole matrix and reads/writes to/from only that portion of the matrix representation on a centralized network drive. This clarifies that each process is doing some algorithmic overhead but is only communicating the data that it needs. While there is certainly room for optimization of the algorithmic part of this operation, the central target we are using is an off the shelf NFS share and by using some other implementation of a POSIX compliant distributed file system (one actually designed for HPC) we could get much higher throughput without changing any of our code. I.e. we are not brushing up against a bottleneck imposed by our poor implementation but rather by the performance of our central data store.

In the course of a DFT calculation, FHI-aims utilizes the following matrices: the overlap matrix $S$, the DFT Hamiltonian $H$, and the density matrix $D$. FHI-aims distributes each matrix across CPU processes and memory using a 2D block-cyclic distribution pattern. On the other hand, in our current TPU implementation which utilizes SUMMA (Scalable Universal Matrix Multiplication Algorithm), our TPU-based solver requires matrices to be distributed across a TPU processor grid as 2D blocks in a \emph{checkerboard} distribution, see~\cite{tpu_algebra} for more details. This poses a practical matrix communication challenge between the CPU-based and TPU-based schemes since their matrix distribution patterns differ in both cyclicity and the number of processors. A simple solution to communicate such matrices between CPU and TPU is to serialize and transfer the respective matrices and deserialize and redistribute them in the desired scheme. Specifically, we utilize available MPI processes on the CPU to serialize (and deserialize) to a network disk using the ELSI IO module and compressed sparse column (CSC) format with no cyclicity. Double precision is used throughout. We note that each process (CPU and TPU) calculates where its data should be within the serialized CSC representation of the whole matrix and reads (writes) to (from) only that portion of the matrix representation on the centralized network drive. That is, our implementation incurs some algorithmic overhead but only communicates the data that is needed.

\begin{figure}[ht]
\includegraphics[width=.45\textwidth]{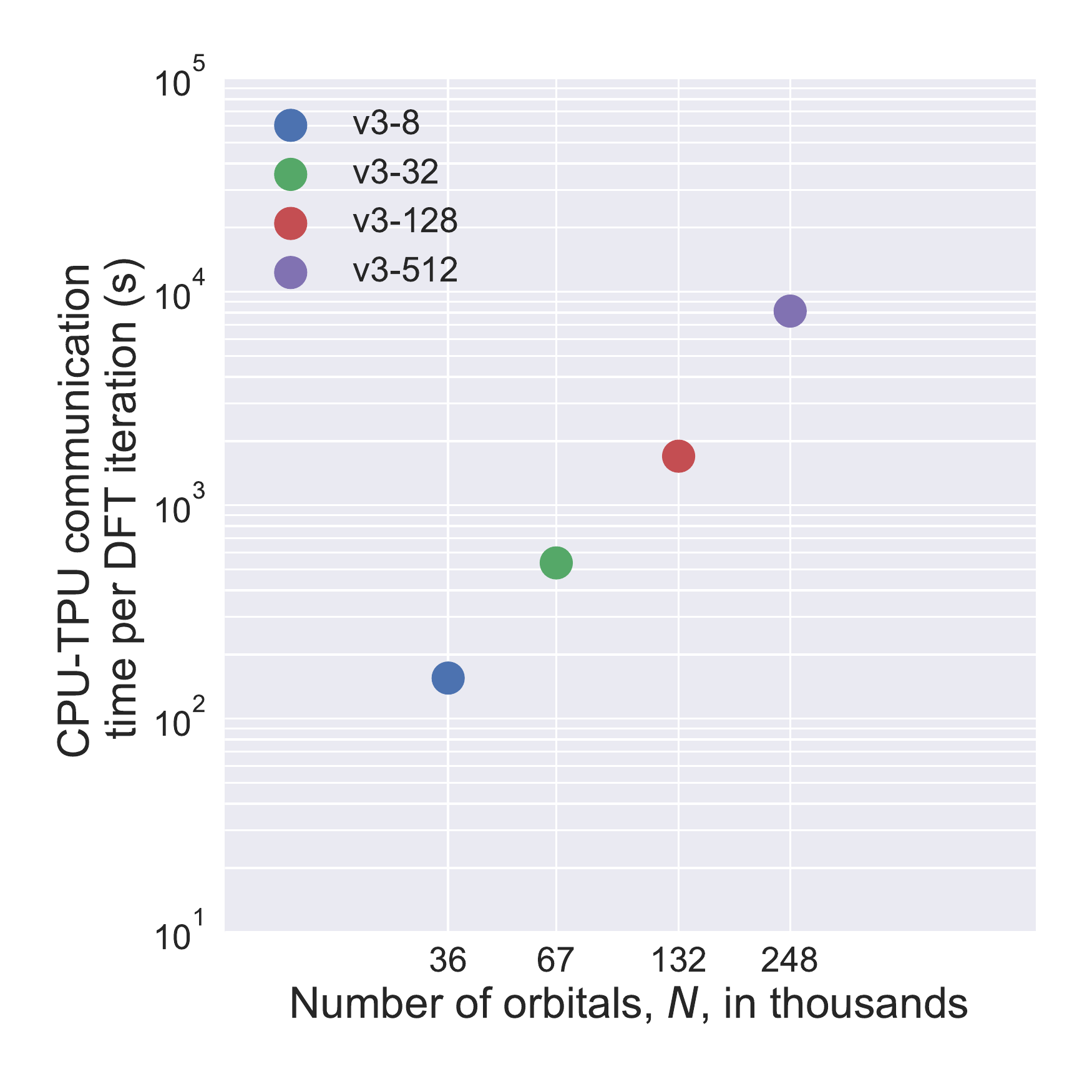}
\centering
    \caption{Total end-to-end CPU-TPU communication time per DFT iteration (excluding the TPU density matrix purification time) incurred in current integration with FHI-aims.} 
\label{fig: io_times}
\end{figure}

This is not the most performant solution, however, it is generalizable, makes use of existing tooling within ELSI, and avoids the complexity of the various distribution patterns. Due to the use of the CSC format, serializing (writing) dense matrices to disk is especially costly and dominates the total CPU-TPU communication time for large system sizes. For transparency, in Fig.~\ref{fig: io_times} we plot the average observed total end-to-end CPU-TPU communication time (excluding the TPU density matrix purification time) incurred in our current implementation. There are, however, several ways to further optimize the current integration. For instance, we are currently using an off-the-shelf network file system (NFS) share, and replacing it with a different implementation of a portable operating system interface (POSIX) compliant distributed file system (one designed for high-performance applications) would result in a much higher throughput and would not require any changes to our code. In addition, further algorithmic optimizations are likely possible.

Within FHI-aims all calculations are performed using the all-electron ``light defaults'' numeric atom-centered basis set~\cite{BLUM20092175}. This results in $5$ basis functions per H atom and $14$ basis functions per O atom in the water cluster calculations. All calculations are non-periodic with open boundary conditions and utilize the PBE~\cite{perdew1996generalized} XC functional. The geometries of water clusters are \edit{directly obtained from Ref.~\cite{ergoscf_waterclusters} which were generated by taking spherical cutouts of varying radii from a large molecular dynamics simulation of bulk water at standard pressure and an average temperature of 300K (further details can be found in Ref.~\cite{rudberg2018ergo}).} 
%The specific geometries used in this work are available as digital materials (Supplementary Data 1). 

In our implementation with FHI-aims, hybrid functionals can also be used without any modification, but simply result in longer DFT Hamiltonian build times on CPUs. Analytical forces are also available from FHI-aims using TPU-computed energy-weighted density matrices, which are also communicated using the above scheme and facilitated using ELSI.\\

\subsection{dynamic precision on smaller water clusters}

\begin{figure}[ht]
\includegraphics[width=.45\textwidth]{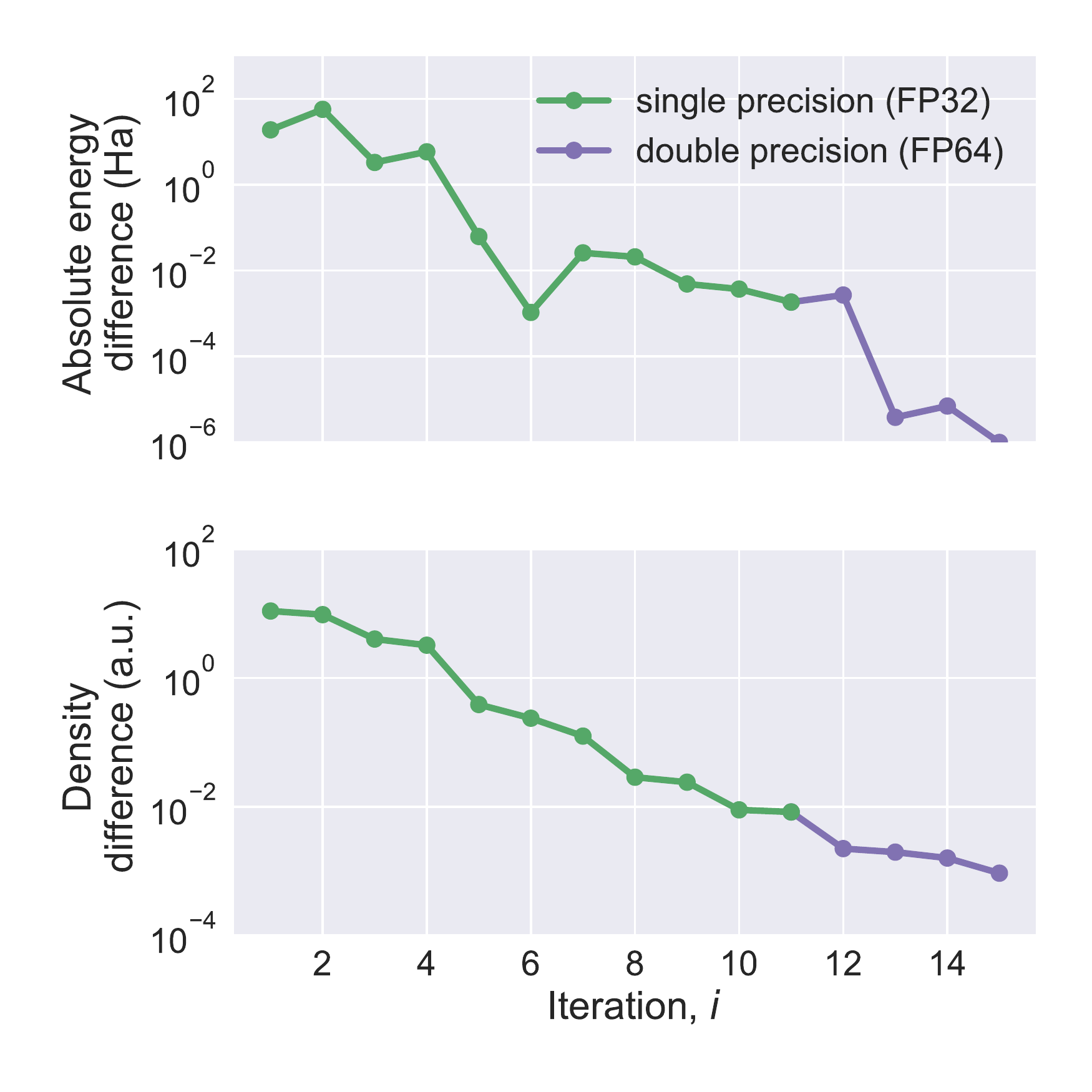}
\centering
\caption{Convergence trajectory of an end-to-end dynamic precision DFT calculation on a (H$_2$O)$_{1481}$ cluster with $N = 35\,544$ orbitals. The absolute total energy differences between subsequent DFT iterations, $i$ and $i-1$, are plotted (top). The corresponding difference in real-space densities within the L$^1$ norm is plotted (bottom).} 
\label{fig:water35544_bifdyn_8_dyn_prec_plot}
\end{figure}

The dynamic precision approach illustrated in Fig.~4 of the main text for 10\,327 water molecules, when applied to smaller systems, allows for a larger part of the computation to be performed in single precision. For instance, an end-to-end converged DFT calculation on (H$_2$O)$_{1481}$ cluster with $N = 35\,544$ orbitals  required 11 iterations in  single  precision  and  4  iterations  in  double  precision, with an overall time of under 5 hours on a single TPU (v3-8) board, see Fig.~\ref{fig:water35544_bifdyn_8_dyn_prec_plot}.

\clearpage

\bibliography{references}

\end{document}